\documentclass[reprint,showpacs,preprintnumbers,amsmath,amssymb,superscriptaddress,bibnotes, apl]{revtex4-1}

\usepackage{graphicx}
\usepackage{dcolumn}
\usepackage{tabularx}
\usepackage{bm}
\usepackage{color}
\makeatletter
\def\Ddots{\mathinner{\mkern1mu\raise\p@	
\vbox{\kern7\p@\hbox{.}}\mkern2mu
\raise4\p@\hbox{.}\mkern2mu\raise7\p@\hbox{.}\mkern1mu}}
\makeatother

\usepackage{lineno}
\begin{document}

\preprint{Preprint}

\title{Sub-cycle temporal evolution of light-induced electron dynamics in {hexagonal 2D materials}}

\author{Christian Heide} 
\email[E-mail: ]{christian.heide@fau.de}
\affiliation{Laser Physics, Department of Physics, Friedrich-Alexander-Universit\"at Erlangen-N\"urnberg (FAU), Staudtstrasse 1, D-91058 Erlangen, Germany}
\author{Tobias Boolakee}
\email[E-mail: ]{tobias.boolakee@fau.de}
\affiliation{Laser Physics, Department of Physics, Friedrich-Alexander-Universit\"at Erlangen-N\"urnberg (FAU), Staudtstrasse 1, D-91058 Erlangen, Germany}
\author{Takuya Higuchi}
\affiliation{Laser Physics, Department of Physics, Friedrich-Alexander-Universit\"at Erlangen-N\"urnberg (FAU), Staudtstrasse 1, D-91058 Erlangen, Germany}
\author{Peter Hommelhoff}
\email[E-mail: ]{peter.hommelhoff@fau.de}
\affiliation{Laser Physics, Department of Physics, Friedrich-Alexander-Universit\"at Erlangen-N\"urnberg (FAU), Staudtstrasse 1, D-91058 Erlangen, Germany}
\date{\today}

\begin{abstract}
	{Two-dimensional materials with hexagonal symmetry such as graphene and transition metal dichalcogenides} are unique materials to study light-field-controlled electron dynamics inside of a solid. Around the $K$-point, the dispersion relation represents an ideal system to study intricately coupled intraband motion and interband (Landau-Zener) transitions driven by the optical field of phase-controlled few-cycle laser pulses. Based on the coupled nature of the intraband and interband processes, we have recently observed in graphene repeated coherent Landau-Zener transitions between valence and conduction band separated by around half an optical period of $\sim$1.3\,fs [Higuchi \textit{et al}., Nature 550, 224 (2017)]. Due to the low temporal symmetry of the applied laser pulse, a residual current density and a net electron polarization are formed. Here we show extended numerical data on the temporal evolution of the conduction band population of 2D materials with hexagonal symmetry during the light-matter interaction, yielding deep insights to attosecond-fast electron dynamics. {In addition, we show that a residual ballistic current density is formed, which strongly increases when a band gap is introduced. Both, the sub-cycle electron dynamics and the resulting residual current are relevant for the fundamental understanding and future applications of strongly driven electrons in two-dimensional materials, including graphene or transition metal dichalcogenide monolayers.}
\end{abstract}
\maketitle

	\begin{figure*}[t!]
	\begin{center}
		\includegraphics[width=14cm]{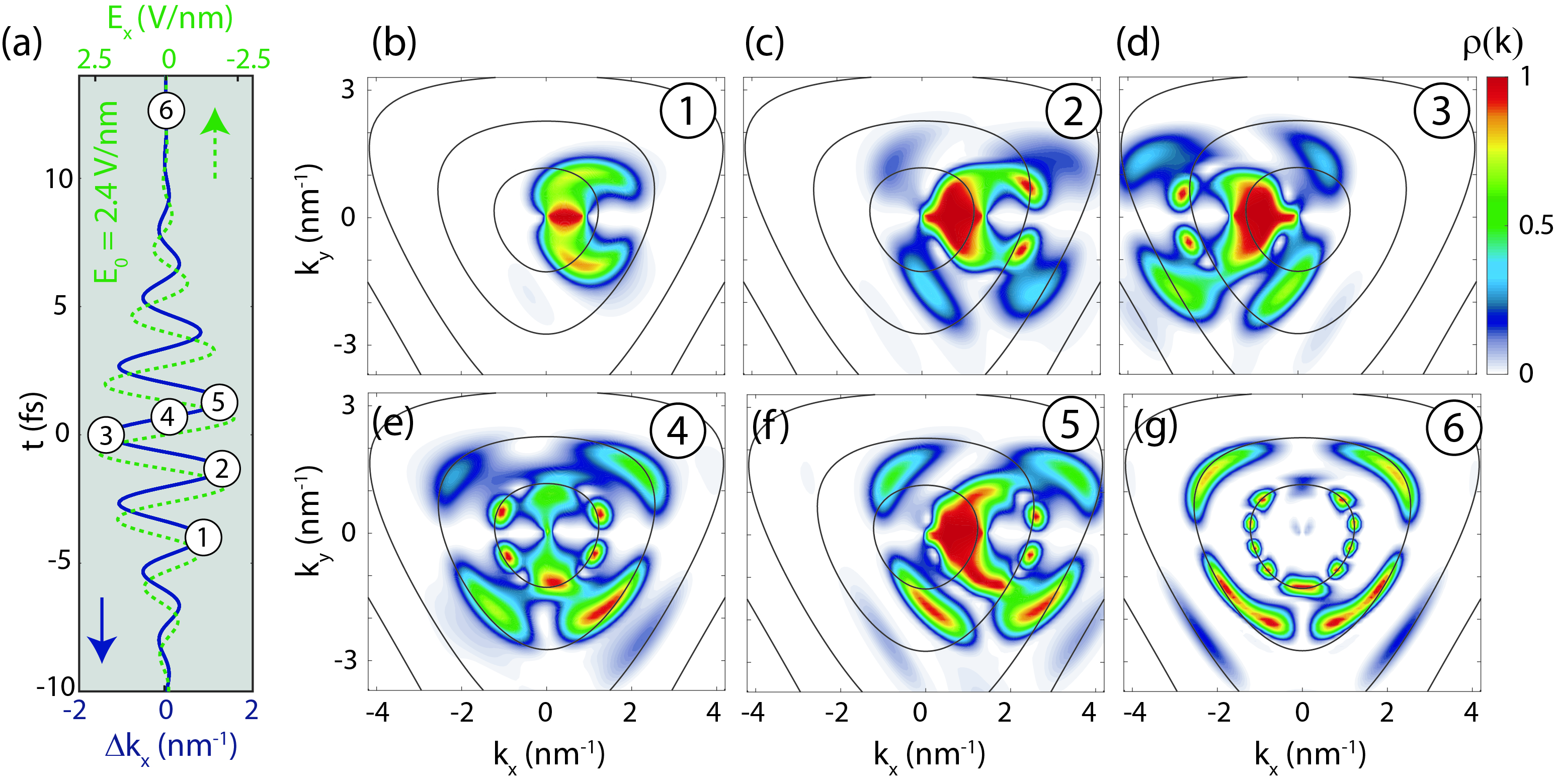}
		\caption{\textbf{Time-dependent conduction band population.} Calculated temporal evolution of the conduction band for an ultrashort $x$-polarized laser pulse with $\tau_\text{p} = 6$\,fs, $\Phi_\text{CEP} = \pi/2$ and $E_0 = 2.4$\,V/nm. \textbf{(a)} The electron trajectory{ of the electron} in the reciprocal space is given by the temporal evolution of $A_x(t)$. For a given electric field waveform (green dashed line), the temporal evolution of the Bloch trajectory $k_\text{x}$ is calculated ({Eq.\,\eqref{Eq. acceleration theorem}} {Eq.\,\eqref{Bloch acceleration equation}}, solid blue line). \textbf{(b)}--\textbf{(g)} Snapshots of the conduction band population during the light-matter interaction are shown. When the vector potential is positive, the conduction band population is shifted to negative $k_\text{x}$-direction and vice versa. During the laser pulse the electron can undergo several Landau-Zener transitions, which may result in a residual asymmetric conduction band population (Fig.~\ref{Fig. 1}\,(b)).}
		\label{Fig. 2}
	\end{center}
\end{figure*}

Recently, the manipulation of electrons at optical frequencies inside of solids using the electric field waveform of laser pulses has found particular interests \cite{Ghimire2014, Krausz2014, Higuchi2017, Kruchinin2018, Reimann2018, Ghimire2019}. Under the presence of an electric field $\textbf{E}(t)$, the change of the electron wave number can be treated semi-classically and is generally described based on the Bloch acceleration theorem \cite{Bloch1928, Kelardeh2014a, Kruchinin2018}: 
\begin{align}
	\dot{\textbf{k}}(t) = -\hbar^{-1}e\textbf{E}(t),
	\label{Eq. acceleration theorem}
\end{align}
where $e>0$ is the absolute value of the elementary charge of an electron. For small electric fields, this change in the electron wave number can be neglected and light-matter interaction is described as a pure interband transition \cite{Meschede2004}. However, when the electric field strengths becomes large (i.e., here several volt per nanometer), the change of the electron wave number (intraband motion) can significantly affect interband transitions \cite{Keldysh1965, Schiffrin2013, Krausz2014, Kelardeh2014a, Ghimire2014, Vampa2015a, Wachter2015, Wismer2016, Chizhowa2016, Higuchi2017, Heide2018, Kruchinin2018, Heide2019, Ghimire2019,OliaeiMotlagh2019a, OliaeiMotlagh2019b}.

A Bloch trajectory for an electron driven with an $x$-polarized few-cycle laser pulse $E_\text{x}(t)~=~E_0~\exp\left(-2\ln(2)(t/\tau_\text{p})^2\right) \cos\left(2\pi f_0 {t}+ \Phi_\text{CEP}\right)$, with a pulse duration of $\tau_\text{p} = 6$~fs, a driving frequency of $f_0$ = 0.375\,PHz and a carrier-envelope phase of $\Phi_\text{CEP} = \pi/2$ is shown in Fig.\,\ref{Fig. 2}\,(a). For a peak electric field strength of $E_0$ = 2.4~V/nm acting on graphene, which is easily reachable with a commercial few cycle laser oscillator, the electron wave number is changed by $\Delta k_\text{x}$ = 1.6\,nm$^{-1}$. We note that this Bloch trajectory is solely governed by the pulse waveform. For materials with a strong interband transition dipole matrix element, such a change in $\mathbf{k}$ may result in Landau-Zener (LZ) transitions between different bands, and the light-matter interaction becomes sensitive the electric field waveform \cite{Landau1932, Zener1932, Stuckelberg1932, Shevchenko2010}. In particular, it has been experimentally and numerically demonstrated that in graphene, a highly inversion-symmetric system with a strong interband coupling at the Dirac point, subsequent LZ transitions can interfere \cite{Ishikawa2010, Ishikawa2013, Kelardeh2014a, Kelardeh2015, Higuchi2017, Heide2018,  OliaeiMotlagh2019b}. Depending on the {symmetry of the applied laser pulse and the }accumulated quantum-mechanical phase between subsequent LZ transitions, a residual current density can be generated. The carrier-envelope phase $\Phi_\text{CEP}$ is used as a characteristic parameter to control the asymmetry of the laser pulse, which is required to generate a residual current density \cite{Franco2008, Higuchi2017}. We note that LZ physics investigated in this study is closely related to molecular ionization physics \cite{Vrakking1996, Shapiro1999, Lezius2001, Lezius2002, Sussman2006}.

Here, we show detailed simulation results of the temporal evolution of the conduction band states in {undoped} monolayer graphene, during the light-matter interaction. For this, we use the nearest-neighbor tight-binding Hamiltonian for a hexagonal lattice \cite{Kelardeh2014a, Higuchi2017, Neto2007}
\begin{align}
	\mathcal{H}({\bf k}(t)) = \left[ \begin{array}{cc} 
		{\Delta/2} & -\varepsilon_\text{h} f \left( {\bf k}(t) \right) \\ 
		-\varepsilon_\text{h} f^* \left( {\bf k}(t) \right) & {-\Delta/2} \end{array} \right],
	\label{Eq. Hamiltonian}
\end{align}
with $f({\bf k}) = \exp \left( i \frac{a k_\text{x}}{\sqrt{3}} \right) + 2 \exp \left( -i \frac{a k_\text{x}}{2\sqrt{3}} \right) \cos \left(\frac{a k_\text{y}}{2} \right)$, $\varepsilon_\text{h}~=~3$\,eV the hopping parameter between nearest neighbor atoms and $a~=~0.246$\,nm the lattice constant of graphene. 
{The underlying crystal structure is a honeycomb crystal structure with two sub-lattices A and B \cite{Neto2007}. $\Delta$ is the band gap between the valence and conduction band at the K-point, which is zero for graphene and nonzero for gapped graphene. The energy of the valence ($\alpha =$ VB) and conduction ($\alpha =$ CB) band states are given as:
\begin{align}
	E_\alpha = \pm \sqrt{(\epsilon_\text{h} |f({\bf k})|)^2 + (\Delta/2)^2},
\end{align}
with $E_\alpha > 0$ for the CB and $E_\alpha < 0$ for VB states.
First we will discuss the lightfield-induced electron dynamic for graphene, i.e., $\Delta = 0$.} The temporal evolution of ${\bf k}(t)$ is described with the Bloch acceleration theorem 
{\begin{align}
\Delta {\bf k}(t) = \hbar^{-1} e {\bf A}(t)
\label{Bloch acceleration equation}
\end{align}}
(cf. Eq.~(1)), {with $\Delta {\bf k}(t) = {\bf k}(t) - {\bf k}_0$}, where ${\bf k}_0$ is the initial wave number and ${\bf A}(t) = \int {\bf E}(t') \text{d}t'$ the vector potential. Using a basis spanned by the Houston functions $\varphi^{(H)}_{\alpha, k_0}$ \cite{Houston1940}, the temporal evolution of the conduction band population is numerically calculated using the Crank-Nicolson algorithm. We assume that any interaction of the electron with other electrons or phonons can be neglected. This assumption is justified when the electron is driven on a timescale faster than the characteristic time constant for electron-electron ($\tau_\text{e,e}\,\approx\,10-80$\,fs) or electron-phonon ($\tau_\text{e,ph}\,\approx\,1-3$\,ps) scattering \cite{Breusing2011, Johannsen2013, Gierz2013, Malic2011, Lui2010}, which is the case for two-cycle laser pulses at 800~nm. 

\begin{figure*}[t!]
	\begin{center}
		\includegraphics[width=14cm]{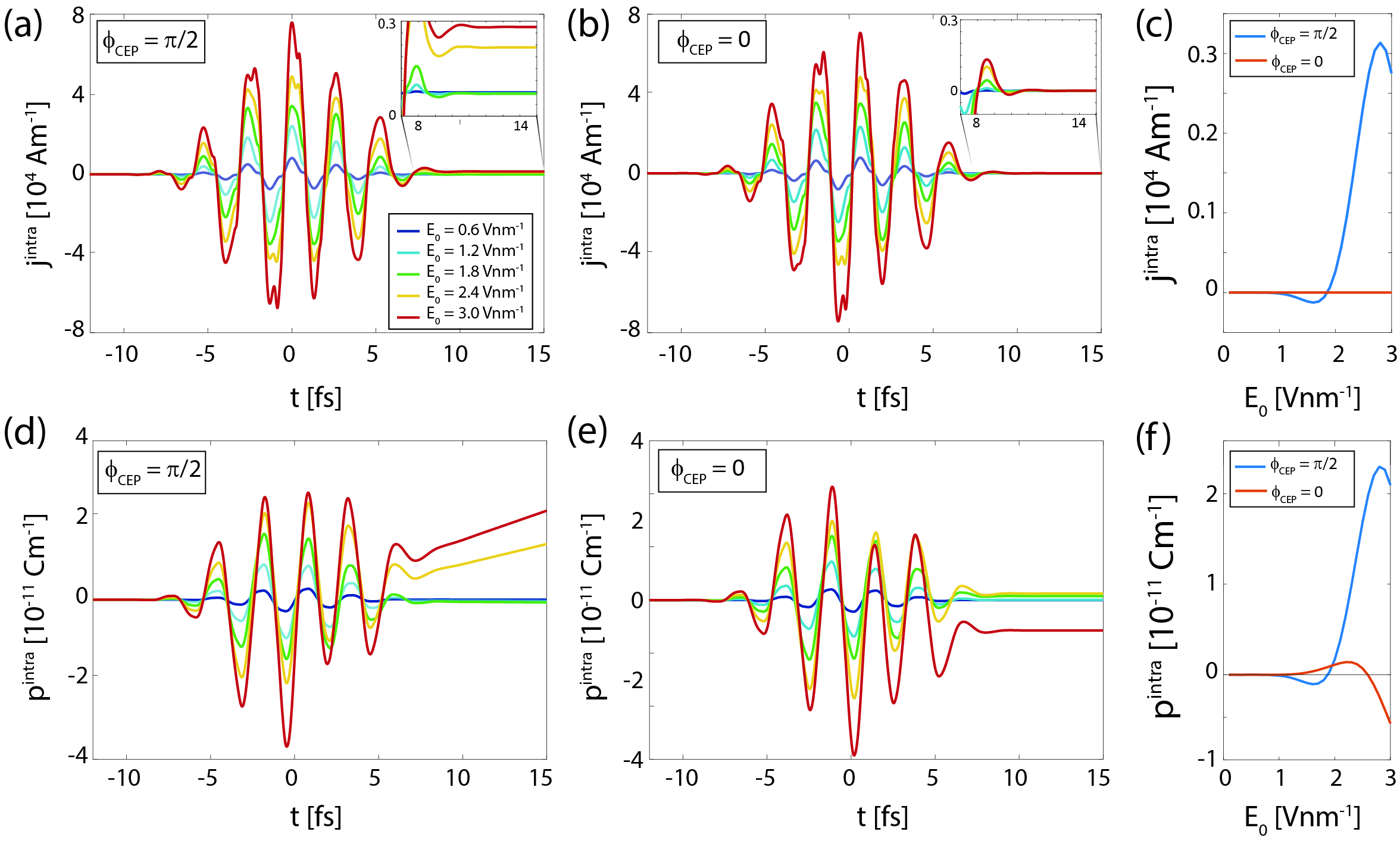}
		\caption{\textbf{Intraband current generation.} Calculated temporal evolution of the intraband current for $\Phi_{\text{CEP}} = \pi/2$ in \textbf{(a)} and $\Phi_{\text{CEP}} = 0$ in \textbf{(b)}, based on Eq.\,\eqref{current density}. The different colors represent different electric field strengths (increasing from 0.6\,V/nm (blue) to 3.0\,V/nm (red). During the laser pulse, the electron follows the negative vector potential $-A_x(t)$ (see Eq.\,\eqref{Bloch acceleration equation}) and an oscillatory intraband current density is found. For $\Phi_\text{CEP} = -\pi/2$, a residual current density is remaining, which is absent for $\Phi_\text{CEP} = \pi$ (see insets). \textbf{(c)} Increasing the electric field strengths results in a non-monotonic increase of the intraband current, with a current reversal around 2\,V/nm for $\Phi_\text{CEP} = \pi/2$ (blue). \textbf{(d)}--\textbf{(f)} Intraband polarization for $\Phi_\text{CEP} = \pi/2$ in (d) and $\Phi_\text{CEP} = 0$ in (e). Similarly to the current, $\Delta p^\text{intra}(E_0)$ reveals a current reversal around 2~V/nm.}
		\label{Fig. 3}
	\end{center}
\end{figure*}

The simulation of the time-dependent conduction band population enables us to track the temporal evolution of the carriers in \textbf{k}-space \textit{during} the interaction with the laser pulse, by projecting the population to the unperturbed conduction band state, as depicted in Fig.\,\ref{Fig. 2} (b)--(g) for $\Phi_\text{CEP} = \pi/2$. {When the vector potential is positive (e.g., Fig.\,\ref{Fig. 2}~(b),~(c),~(f)), the electron gains momentum towards positive $k_\text{x}$-values, whereas for negative ${\bf A}(t)$, such as in Fig.\,\ref{Fig. 2}~(d), the electron gains momentum towards negative $k_\text{x}$-values.} In particular, when the vector potential is so strong that $\Delta \mathbf{k} = \hbar^{-1} e \mathbf{A}(t) > \mathbf{k}_0$, the driven electron passes by the Dirac point twice per optical cycle and the electron can transition from one to the other band. {For ${\bf A}(t) = 0$, the electron is back at its original wave number (e.g., Fig.\,\ref{Fig. 2} (e), (g)).}\\
During the laser pulse, an asymmetric CB and VB population distribution $\rho^\text{(CB)}$ is generated, which is the source of the intraband current
\begin{align}
	j^\text{intra}(t) =  2 g_\text{s} e \int_\text{BZ}\rho^\text{(CB)}(\mathbf{k}, t) v_\text{x}(\mathbf{k})  \frac{d\mathbf{k}}{(2\pi)^2},
	\label{current density}
\end{align}
with $v_\text{x}(\mathbf{k}) = \frac{\partial\varepsilon(\mathbf{k})}{\partial k_\text{x}}$ the group velocity. The factor 2 accounts for electrons and holes and $g_\text{s} = 2$ for two kinds of spins. The integral is taken over the Brillouin zone of graphene. Based on Eq.\,\eqref{current density} the current density is calculated for different electric field strengths and carrier-envelope phases. Figure \ref{Fig. 3}\,(a) shows the temporal evolution of $j^\text{intra}(t)$ for $\Phi_{\text{CEP}} = \pi/2$ and various electric field strengths from $E_0 = 0.1 ... 3$\,V/nm. Increasing the peak electric field strength results in a longer electron Bloch trajectory and in a larger intraband current. During the laser pulse, the conduction band population and thus $j^\text{intra}$ follows the vector potential. For high electric field strengths (red lines), high harmonics of the intraband current are generated, which can be directly seen in the strong deviation from a {sinusodial} {sinusoidal} behaviour of the largest instantaneous currents \cite{McDonald2015, Chizhowa2017, Kim2019}. After the laser pulse is gone, a residual current density is found, which results in a ballistic current (Fig.~\ref{Fig. 3}\,(c)). Strikingly, the direction of this ballistic current reverses at around 2\,V/nm. We have shown in \cite{Higuchi2017, Heide2019} that the current reversal can be measured and that it is determined by the accumulated phase during subsequent LZ transitions. Note that flipping the phase by $\pi$ to $\Phi_{\text{CEP}} = \pi/2$ mirrors the vector potential and thus the current direction \cite{Higuchi2017,OliaeiMotlagh2019a}. For $\Phi_{\text{CEP}} = 0$ and $\Phi_{\text{CEP}} = \pi$, a large intraband current density during the laser pulse is found, however no residual current (Fig.~\ref{Fig. 3}\,(a), (b)).\\

The intraband current density leads to a net transferred charge, also known as intraband polarization
\begin{align}
	P^\text{intra}(t) = \int_{-\infty}^t j^\text{intra}(t')\text{d}t'.
\end{align}
During the lifetime of the ballistic current, generated for $\Phi_{\text{CEP}} = \pm \pi/2$, the electrons (in CB) and holes (in VB) are separated and $P^\text{intra}(t)$ increases linearly as a function of the time {(see Fig.\,~\ref{Fig. 3}\,(d))}. This charge displacement will become constant when the asymmetry of CB-population isotropizes and the ballistic current decays (not shown, time scale $\sim$10\,fs). Subsequently, electron-hole recombination equilibrate{s} the system on a time scale of 100\,fs to 1\,ps \cite{Breusing2011, Johannsen2013, Brida2013, Gierz2013, Malic2011, Lui2010}.\\
Due to the nonlinear dependence of the Landau-Zener process on the electric field strengths \cite{Higuchi2017, Shevchenko2010}, even for the case of $\Phi_{\text{CEP}} = 0$ and $\pi$ where the residual current is zero, the temporal evolution of the current density has a nonzero area, which results in a nonzero transferred charge density, see Fig.\,\ref{Fig. 3}\,{(d)}--(f). Similar to the current, the polarization shows a reversal of the direction of net charge transport for $\Phi_{\text{CEP}} = \pi/2$ and for $\Phi_{\text{CEP}} = 0$, depicted in Fig.\,\ref{Fig. 3}\,(f).\\

	\begin{figure*}[t!]
	\begin{center}
		\includegraphics[width=14cm]{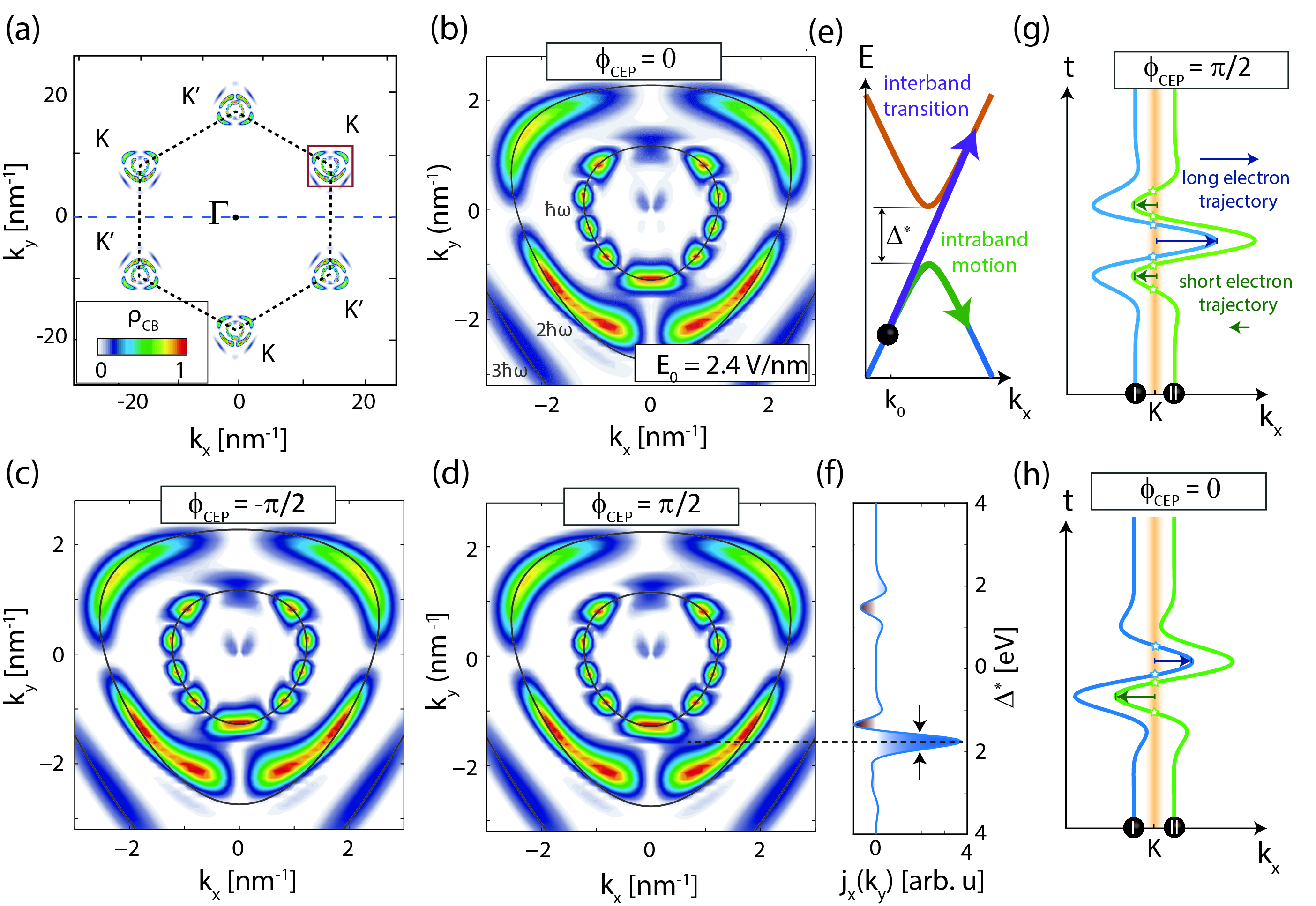}
		\caption{\textbf{Residual conduction band population.} 
			\textbf{(a)} TDSE model simulation results of the conduction band population $\rho_{\text{CB}}$ over the whole first Brillouin zone after interaction with ultrashort laser pulses is shown. The parameters for the $x$-polarized laser pulse are $\tau_\text{p} = 6$\,fs, $\Phi_\text{CEP} = \pi/2$ and $E_0 = 2.4$\,V/nm, $\omega_0 = 2\pi \cdot 0.375$\,PHz. For this electric field strength and driving frequency, the CB population is found near the $K$ and $K'$ point. The conduction band population distribution $\rho_{\text{CB}}(\textbf{k})$ for $K$ and $K'$ points are mirror symmetric with respect to the $k_\text{x}$-axis through the $\Gamma$-point (blue dashed line). 
			\textbf{(b)}--\textbf{(d)} Magnified residual conduction band population $\rho_{\text{CB}}$ in the vicinity of the $K$ point after excitation with $\Phi_\text{CEP} = \pi/2$ (b), $\Phi_\text{CEP} = -\pi/2$ (d) and $\Phi_\text{CEP} = 0$ (e) is plotted. $\rho_{\text{CB}}$ = 1 means that the electron is excited from VB to CB. Solid curves indicate the resonances, where the energy difference between the valence and conduction band corresponds to the photon energy $\hbar\omega$, $2\hbar\omega$ or $3\hbar\omega$. Depending on $\Phi_\text{CEP}$, an off-resonant CB population is found (red circle). {\textbf{(e)} Current density for various initial $k_y$. The maximal current density is found for electrons with an initial wave number of $k_\text{y} = 1.6$nm$^{-1}$, corresponding to a effective band gap $\Delta^\ast$ = 1.75\,eV.
				\textbf{(f)} The electron starting at $k_0$, driven by an electric field can undergo intraband motion or interband transition near the $K$-point within a half optical cycle of the laser field. }
			\textbf{(g, h)} Electron trajectory for $\Phi_\text{CEP} = \pi/2$ in (g) and $\Phi_\text{CEP} = 0$ in (h) for an electron with an initial negative and positive wave number. The orange line represents the region, where the electron can undergo a LZ-transition. For illustration we use a single cycle laser pulse.}
		\label{Fig. 1}
	\end{center}
\end{figure*}

\begin{figure*}[t!]
	\begin{center}
		\includegraphics[width=14cm]{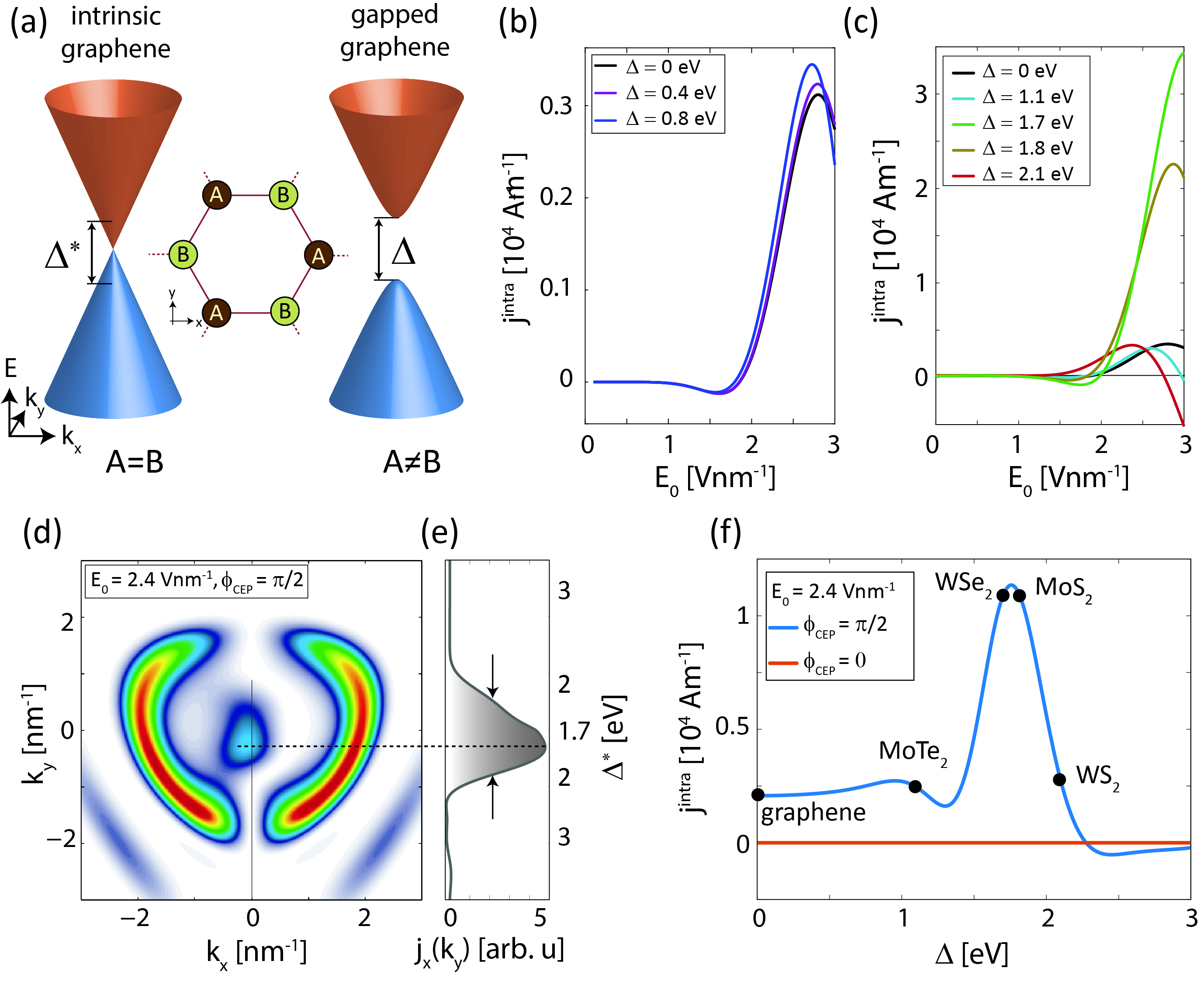}
		\caption{\textbf{Residual current in gapped graphene.} \textbf{(a)} Schematic illustration of the band structure of graphene without a band gap and gapped graphene with a band gap $\Delta$. Depending on the initial $k_y$, the electron is driven along a slice with an apparent band gap $\Delta^\ast$ (see Fig.\,\ref{Fig. 1}\,(e)).
			\textbf{(b)} $j^\text{intra}$ versus $E_0$ for a band gap smaller than the photon energy. The current barely depends on $\Delta$. 
			\textbf{(c)} For a band gap comparable or larger to the photon energy, $j^\text{intra}$ strongly depends on $\Delta$. 
			\textbf{(d)} Residual conduction band population for around the $K$-point of gapped graphene with a band gap of $\Delta = 1.7$. 
			\textbf{(e)} Intraband current as a function of $k_\text{y}$ and corresponding $\Delta^\ast$. A strong population asymmetry is found in the vicinity of the $K$-point. 
			\textbf{(f)} Total $j^\text{intra}$ as a function of $\Delta$ at a field strength of $E_0$ = 3 V/nm and a carrier envelope phase of $\pi/2$. For $\Delta = 1.7$\,eV the CEP-dependent current is maximized. Here, the electron undergoes efficiently intraband motion and interband transitions, resulting into a residual asymmetric current density (see inset). For a CEP of $0$ or $\pi$ no residual current is found. $j^\text{intra}$ for $\Delta = 1.7$\,eV is 40 times larger than for ungapped graphene. The band gap of graphene ($\Delta$ = 0\,eV), MoTe$_2$ ($\Delta$ = 1.1\,eV), WSe$_2$ ($\Delta$ = 1.7\,eV),  MoS$_2$ ($\Delta$ = 1.8\,eV) and WS$_2$ ($\Delta$ = 2.1\,eV) is highlighted.}
		\label{Fig. 4}
	\end{center}
\end{figure*}

To understand the peculiar residual conduction band asymmetry, we focus on the final conduction band population $\rho_{\text{CB}}$. Figure \ref{Fig. 1}\,(a) shows $\rho_{\text{CB}}$ of the first Brillouin zone, i.e., \textit{after} the laser excitation pulse has passed, with the $\Gamma$-point at (0, 0), $K = \frac{2\pi}{3a}\left(1,\frac{1}{\sqrt{3}}\right)$ and $K' = \frac{2\pi}{3a}\left(1,-\frac{1}{\sqrt{3}}\right)$ \cite{Neto2007}. To drive an electron from $K$ to $K'$ (using $y$-polarized light) or to another $K$ point (using $x$-polarized light), an electric field strength of $E_{0, \text{y}} = 15.2$\,V/nm and $E_{0, \text{x}} = 26.4$\,V/nm at 800\,nm is required, respectively. For the laser parameters considered here, we do not drive such inter-vall{e}y processes, thus we can focus on the region around a single $K$ point, magnified in Fig.\,\ref{Fig. 1}\,(b)--(d) for different carrier envelope phases. The solid black lines are resonance lines, corresponding to an energy difference between VB and CB of $n\hbar\omega$ with $\hbar\omega = 1.55\,$eV and $n$~=~1,~2,~... . At these resonances one-photon ($n$ = 1) or multi-photon absorption ($n$~=~2,~3,~...) dominates the light-matter interaction. In the case of $\Phi_\text{CEP} = 0$, illustrated in Fig.\,\ref{Fig. 1}\,(b), the conduction band population distribution is fully mirror symmetric with respect to $k_\text{y} = 0$ and thus no residual asymmetric conduction band population distribution is found. By changing the carrier-envelope phase from $0$ to $\pm\pi/2$, as shown in Fig.\,\ref{Fig. 1}\,(c) and (d), an off-resonant, asymmetric conduction band population is found \cite{Higuchi2017, OliaeiMotlagh2019a}. 
{The hot-spot of this asymmetric population emerges off-resonant, between the one-photon and two-photon absorption resonances. Here, the electron is driven on a trajectory in $x$-direction through the band structure, with an apparent band gap of $\Delta^\ast$, as depicted in Fig.\,\ref{Fig. 1}\,(e). Within one optical cycle, the electron can undergo first an intraband motion followed by an interband transition or vice versa. Both quantum-pathways have the same initial and final wave number and may interfere. This so-called Landau-Zener-Stückelberg (LZS) interference is determined by two criteria, first, the splitting ratio of interband transition and interband motion, which is governed by $\Delta^\ast$, and second, the accumulated phase, which is given by the symmetry of the waveform, i.e., the CEP.\\
Figure\,\ref{Fig. 1}\,(f) shows the current density integrated along $k_\text{x}$ only. The width of this current distribution along $k_\text{y}$ is given by the slope of the dispersion relation $\partial\varepsilon(\mathbf{k})/\partial k_\text{y}$ around $\Delta^\ast \approx 1.75$\,eV (see black arrows). At these off-resonances, where $\Delta^\ast \approx 1.75$\,eV, maximal CEP-dependent conduction band population is obtained. The direction of the residual current density is given by the phase accumulated during the pulse.\\
In Fig.\,\ref{Fig. 1}\,(g) and Fig.\,\ref{Fig. 1}\,(h) two electron trajectories are shown: one for an electron starting at initial negative wave number (I) and one with positive initial wave number (II). For illustration we use a single laser pulse with a carrier envelope phase of $\Phi_{\text{CEP}} = \pi/2$ (Fig.\,\ref{Fig. 1}\,(g)) and $\Phi_{\text{CEP}} = 0$ (Fig.\,\ref{Fig. 1}\,(h)). In case of (I), the electron  experiences two LZ transition events per pulse (blue stars), with a long electron trajectory after the first LZ transition. In contrast, in case of (II), the electron undergoes four LZ transitions (green stars) with two short electron trajectories. The quantum mechanical phase accumulation, which determines the final LZS interference condition, is different for two initial wave numbers. As a result, an asymmetric conduction population emerges. In case of $\Phi_\text{CEP} = 0$, shown in Fig.\,\ref{Fig. 1}\,(h), both trajectories undergo mirrored trajectories with two LZ transitions (see green and blue stars). Thus, the total accumulated phase for cases (I) and (II) are equal and no residual asymmetry is generated.}

{So far we have discussed graphene with $\Delta = 0$, where both sub-lattices A and B consists of indistinguishable carbon atoms. When A and B are different, the inversion symmetry is broken and a band gap at $K$ and $K'$ is opened (Fig.\,\ref{Fig. 4}\,(a)). Gapped graphene and monolayer transition metal dichalcogenides (TMDCs) are such materials, exhibiting a direct band gap at $K$ and $K'$ \cite{Mak2010, Splendiani2010, Wang2012, Kormanyos2015, Kibis2017, Fang2018, OliaeiMotlagh2019b}. Note that both material classes can be approximated by the same Hamiltonian (Eq.\,\eqref{Eq. Hamiltonian}, \cite{OliaeiMotlagh2019b}). The band structure near the $K$-point is schematically illustrated in Fig.\,\ref{Fig. 4}\,(a). When $\Delta$ is smaller than the photon energy, e.g., $\Delta = 0.4$\,eV (purple line, Fig.\,\ref{Fig. 4}\,(b)) or $\Delta = 0.8$\,eV, (blue line, Fig.\,\ref{Fig. 4}\,(b)) the residual current density $j^\text{intra}$ barely depends on band gap. In contrast, when $\Delta$ is comparable to the photon energy, the residual current density strongly depends on the value of the band gap. The underlying residual conduction band population for $\Delta = 1.75$\,eV is shown in Fig.\,\ref{Fig. 4}\,(c). In contrast to $\Delta = 0$ (Fig.\,\ref{Fig. 4}\,(d)), one photon absorption is strongly suppressed, and an asymmetric conduction band population is found near the $K$ point. Due to the hyperbolic band structure, the slope of the dispersion is decreased near the $K$-point, resulting in a larger number of electrons contributing to Landau-Zener-Stückelberg interference \cite{Shevchenko2010}. As result, the width of the current distribution along $k_\text{y}$, shown in Fig.\,\ref{Fig. 4}\,(e), is increased (compare to Fig.\,\ref{Fig. 1}\,(c)). Note that a reduction of the group velocity is taken into account. Fixing $E_\text{0} = 2.4$\,V/nm and sweeping $\Delta$ shows a pronounced peak for $\Delta = 1.75$\,eV. A smaller band gap results in fewer electrons contributing due to an increase in $\partial\varepsilon(\mathbf{k})/\partial k_\text{x}$ around $\Delta^\ast = 1.75$\,eV and a larger band gap favors intraband transitions and thus leads to less efficient Landau-Zener-Stückelberg interference. For the laser parameters considered here, the maximal CEP-dependent current is found for $\Delta$ = 1.75\,eV.
Such a band gap can be found in TMDCs, such as WSe$_2$ with a band gap of $\Delta = 1.7$\,eV and MoS$_2$ ($\Delta$ = 1.8\,eV). For smaller driving frequencies, WTe$_2$ with $\Delta$ = 1.1\,eV becomes more efficient and for larger driving frequencies WS$_2$ ($\Delta$ = 2.1\,eV) \cite{Mak2010, Kumar2012}.}

So far, only the residual currents in graphene have been directly measured. Our results discussed here are based on numerical simulations. To measure the sub-optical cycle evolution of the light-induced electron dynamics at optical frequencies directly, time- and angle-resolved photoemission spectroscopy (trARPES) is required, which becomes nowadays feasible using attosecond pulses \cite{Sie2019}. Recently, such electron dynamics, driven with terahertz pulses at topological surface states in Bi$_2$Te$_3$, have been measured using UV pulses as a probe \cite{Reimann2018}. However, direct measurements of current \textit{driven} at optical frequencies are still missing. This time regime is in particular important to study fundamental processes such as electronic decoherence, resulting from dephasing and electron-electron scattering on the femtosecond timescale. Alternatively to trARPES, indirect measurements, such as the generation of high-harmonics \cite{McDonald2015, Chizhova2017, Yoshikawa2017, Hafez2018, Kim2019, Han2019} or a residual conducing current \cite{Higuchi2017, Heide2018, Heide2019, Wu2019} can reveal information about the electron dynamics. Whereas the first one is sensitive to the inter- and intraband electron dynamics during the laser pulse, the latter is sensitive to the residual current density, w{h}ich is a result of coherent quantum path interference of different Bloch trajectories \cite{Shevchenko2010, Higuchi2017}.\\

To summarize, we have demonstrated that {two-dimensional materials with a hexagonal symmetry} are highly interesting for the investigation of light-field driven electron dynamics. The strong interband coupling allows driving charges fully coherently from one to the other band, resulting in residual population distribution.  Here, the electric field waveform is used to imprint information encoded in the light to the population distribution. {We found that the residual waveform-sensitive current is maximized for materials with a band gap of around 1.75\,eV.} These results will be interesting for the investigation of the role of electronic decoherence, which is crucial for petahertz light-field information technology. In addition, the concept of light-field driven electrons has recently found attraction to investigate topological effects in gapped materials such as TMDCs using circular polarized driving pulses \cite{Kelardeh2016, Nematollahi2019, OliaeiMotlagh2019}{, which might become particularly interesting for Coulomb engineering of band gap materials \cite{Raja2017}.}

\section*{Acknowledgments}
This work has been supported in part by the European Research Council (Consolidator Grant “NearFieldAtto”), Deutsche Forschungsgemeinschaft (Sonderforschungsbereich 953 “Synthetic Carbon Allotropes”, project 182849149) and the PETACom project financed by Future and Emerging Technologies Open H2020 program. C. H. is part of the Max Planck School of Photonics supported by BMBF, Max Planck Society, and Fraunhofer Society. P. H. greatfully acknowledges a Fellowship from Max Planck Institut of the Sciene of Light (MPL).

\providecommand{\newblock}{}

\end{document}